\documentclass[aps,prl,reprint,superscriptaddress,twocolumn,showpacs,amsmath,amssymb, amsfonts,10pt]{revtex4-1}

\usepackage{graphicx}
\usepackage{dcolumn}
\usepackage{bm}
\usepackage{color}
\usepackage{braket}
\usepackage{cleveref}
\usepackage{lineno}

\crefname{equation}{equation}{equations}
\Crefname{equation}{Equation}{Equations}
\crefrangelabelformat{equation}{(#3#1#4--#5#2#6)}
\crefmultiformat{equation}{(#2#1#3}{, #2#1#3)}{#2#1#3}{#2#1#3}
\Crefmultiformat{equation}{Equations (#2#1#3}{, #2#1#3)}{#2#1#3}{#2#1#3}

\begin{document}

\title{Spin-Holstein models in trapped-ion systems}
\author{J. Kn\"orzer}
\email{johannes.knoerzer@eth-its.ethz.ch}
\affiliation{Max-Planck-Institute of Quantum Optics, Hans-Kopfermann-Straße 1, D-85748 Garching, Germany}
\affiliation{Munich Center for Quantum Science and Technology (MCQST), Schellingstr. 4, D-80799 M{\"u}nchen, Germany}
\author{T. Shi}
\email{tshi@itp.ac.cn}
\affiliation{CAS Key Laboratory of Theoretical Physics, Institute of Theoretical Physics, Chinese Academy of Sciences, P.O. Box 2735, Beijing 100190, China}
\affiliation{CAS Center for Excellence in Topological Quantum Computation, University of Chinese Academy of Sciences, Beijing 100049, China}
\author{E. Demler}
\affiliation{Department of Physics, Harvard University, Cambridge, MA 02138, USA}
\affiliation{Institute for Theoretical Physics, ETH Zurich, 8093 Zurich, Switzerland}
\author{J. I. Cirac}
\affiliation{Max-Planck-Institute of Quantum Optics, Hans-Kopfermann-Straße 1, D-85748 Garching, Germany}
\affiliation{Munich Center for Quantum Science and Technology (MCQST), Schellingstr. 4, D-80799 M{\"u}nchen, Germany}

\date{\today}
\begin{abstract}
    In this work, we highlight how trapped-ion quantum systems can be used to study generalized Holstein models, and benchmark expensive numerical calculations.
    We study a particular \textit{spin}-Holstein model that can be implemented with arrays of ions confined by individual microtraps, and that is closely related to the Holstein model of condensed matter physics, used to describe electron-phonon interactions.
    In contrast to earlier proposals, we focus on simulating many-electron systems and inspect the competition between charge-density wave order, fermion pairing and phase separation.
    In our numerical study, we employ a combination of complementary approaches, based on non-Gaussian variational ansatz states and matrix product states, respectively.
    We demonstrate that this hybrid approach outperforms standard density-matrix renormalization group calculations.
\end{abstract}
\maketitle

Electron-phonon interactions lie at the heart of several phenomena in condensed matter physics, including Cooper pairing \cite{cooper56} and the formation of polarons \cite{frohlich54}.
Generally, the low-energy excitations of electrons in solids are modified by their coupling to lattice vibrations, which alters their transport and thermodynamic behaviour.
Often simplified toy models can be employed to study those essential properties.
As a complementary approach to traditional solid-state methods, quantum simulations utilize the rich toolbox of atomic physics to provide a characterization of equilibrium and dynamical properties of paradigmatic quantum many-body models.

The Holstein model is one such paradigmatic model that features a local coupling between the electron density and optical phonons on a lattice \cite{holstein59}.
Despite its apparent simplicity, it hosts rich physics, giving rise to superconducting (SC) phases, charge-density wave (CDW) order and phase separation (PS) at strong coupling \cite{scalettar89,marsiglio90}.
Yet, notwithstanding recent progress, its numerical treatment is often costly, especially when interactions become increasingly strong or of long-range character.
As a tantalizing prospect, quantum simulators may help to gain new insights into the underlying physical mechanisms, and potential implementations include trapped ions \cite{cirac12,blatt12}, hybrid atom-ion systems \cite{bissbort13}, cold atoms \cite{cuadra18} and quantum dots \cite{utso21}.
In trapped ions, their spin and motional degrees of freedom can be harnessed to realize a quantum-optics analogue of the electron-phonon system \cite{porras04,deng05,stojanovic12,negretti20}, which enables access to a variety of system observables.
Moreover, their key parameters may be tuned \textit{in-situ} to explore different regions of the phase diagram.
Currently available setups may thus be utilized to improve and benchmark analogue quantum simulators against state-of-the-art numerical methods.
This paves a way towards the quantum simulation of even more complex electron-phonon models that could be implemented using trapped-ion setups.

In this Letter, we theoretically investigate such trapped-ion systems and derive an effective model that contains strong and highly non-local interactions between effective spins and lattice phonons.
We highlight its similarities and differences with the Holstein model and develop a powerful numerical toolbox to thoroughly characterize its ground-state properties.
Our numerical method combines density matrix renormalization group (DMRG) calculations \cite{white92} and computations based on non-Gaussian variational ansatz states (NGS) \cite{shi17,shi20}.
This hybrid approach is shown to particularly excel at studying the quantum many-body system at large spin-phonon couplings and large phonon numbers.
We define spin and phonon observables motivated by the physics of the Holstein model and study their characteristics.
Using these observables, we identify SC and CDW phases and their relation to the ion-trap parameters, thus demonstrating the rich Holstein-like physics of the trapped-ion system.
Finite-temperature and finite-size calculations show that our results can be expected to be robust against thermal excitations in state-of-the-art setups.

\begin{figure}[t!]
\includegraphics[width=0.99\columnwidth]{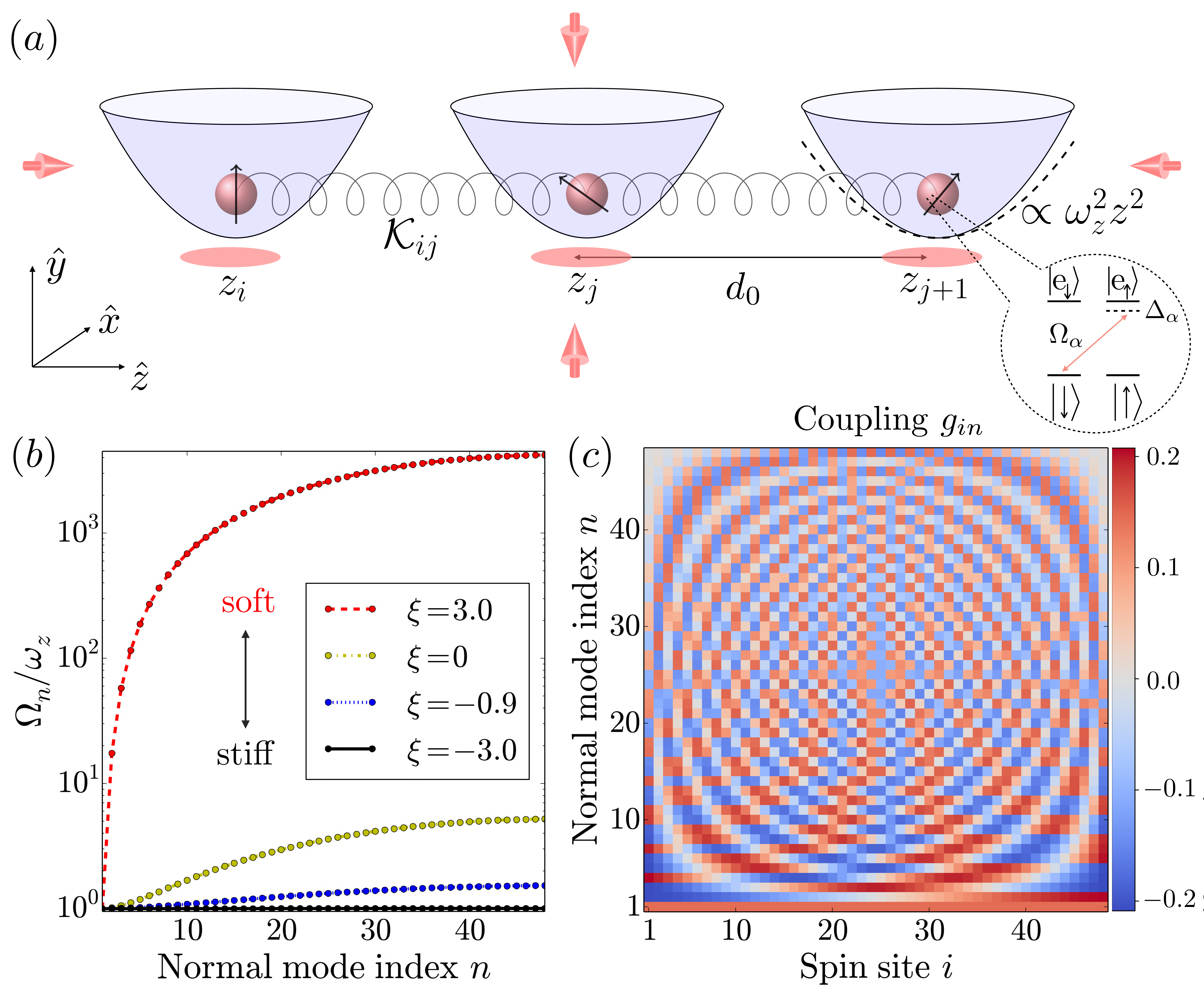}
\caption{\label{fig:sketch}Schematic illustration of setup.
(\textit{a})
Trapped-ion chain subject to three counter-propagating laser beams.
The microtraps are aligned along the $\hat z$ direction at a distance $d_0$.
Ions are coupled to each other via their mutual Coulomb interaction, indicated by springs.
The inset shows an exemplary level scheme with four internal states $\ket{\uparrow}, \ket{\downarrow}, \ket{\mathrm{e}_\downarrow}, \ket{\mathrm{e}_\uparrow}$, and a $\sigma^+$ transition with laser parameters $\Omega_\alpha$ and $\Delta_\alpha$.
(\textit{b}) Normal-mode frequencies $\Omega_n/\omega_z$ for different values of $\xi$.
$\omega_z$ is fixed while $d_0$ is varied.
(\textit{c}) Coupling $g_{in}$ for $\xi = 0$ exemplarily shows long-range interactions between spins and phonons.
}
\end{figure}

\textit{Setup and model.}\textemdash
We consider a physical system of $N$ ions with mass $m$, each confined to a harmonic microtrap to guarantee an equidistant spacing of ions.
All of the ions' equilibrium positions are assumed to be aligned along the $\hat z$ axis at a nearest-neighbour distance $d_0$, see Fig.~\ref{fig:sketch}(a).
In a laser beam configuration which hosts three standing waves along the $\hat x$, $\hat y$ and $\hat z$ axes, light that is off-resonant with chosen hyperfine-state transitions of the ions can be harnessed to introduce a coupling between the motional and spin degrees of freedom of all ions \cite{porras04,deng05}.
We assume large transverse trap frequencies and eliminate the motional degrees of freedom along $\hat{x}$ and $\hat{y}$ via a polaron transformation.
As a result, pseudospins at distance $r$ become effectively coupled through an effective dipolar interaction $J/r^{3}$ at strength $J$.
As outlined in more detail in the supplemental material \cite{SM} we obtain an effective description of our system which takes the form
\begin{equation}
H_{\mathrm{eff}}=\sum_{n}\Omega _{n}a_{n}^{\dagger }a_{n}+\sum_{\substack{ i\neq j,  \\ \alpha =x,y}}\frac{J}{|i-j|^{3}}\sigma _{i}^{\alpha }\sigma_{j}^{\alpha }+H_{\mathrm{int}}  \label{eq:hamilt1}
\end{equation}
where $a_{n=1,...,N}$ are annihilation operators of the $N$ collective phonon normal modes with frequencies $\Omega _{n}$ (see Fig.~\ref{fig:sketch}(b)), $\sigma_i^\alpha$ denotes the Pauli matrix associated with the internal spin states $\ket{\uparrow}$ and $\ket{\downarrow}$ at site $i$ and direction $\alpha$.
In terms of the mode expansion $r_{i}=\sum_{i}g_{in}(a_{n}+a_{n}^{\dagger })$ the interaction $H_{\mathrm{int}}=-F_{z}\sum_{i}r_{i}(1+\sigma_{i}^{z})$ of spins and local longitudinal phonons becomes
\begin{equation}\label{eq:hInt}
H_{\mathrm{int}}=-F_{z}\sum_{i,n}g_{in}(a_{n}+a_{n}^{\dagger })(1+\sigma
_{i}^{z}),
\end{equation}
where $g_{in}$ describes the non-local coupling between phonon normal modes and spins (see Fig.~\ref{fig:sketch}(c)).
Here we have made the Lamb-Dicke approximation, which can be justified in experiment if the light-induced coupling between internal spin states and motional states of the ions is sufficiently small.

Our effective model in Eq.~\eqref{eq:hamilt1} contains several key parameters that determine its behaviour.
In the following, we set $\omega_z/J = 1$ for all microtraps, and focus on the rich physics left to explore with the remaining free parameters.
In particular, the system can now be described by (\textit{i}) the spin-phonon coupling $F_z$ and (\textit{ii}) the ion trap stiffness $\beta = e^2/(m\omega_z^2 d_0^3)$ along the $\hat z$ direction.
Throughout this work, we will use $\xi = \log \beta$.
The limit $\xi \lesssim -1$ ($\xi \gtrsim 1$) is usually referred to as the stiff (soft) limit, in which the phonon dispersion is weak (strong) (see Fig.~\ref{fig:sketch}(b)).
The ion-trap setup allows us to switch between the adiabatic (small phonon frequency) and diabatic (large phonon frequency) regimes of the spin-Holstein model \eqref{eq:hamilt1}.

\textit{Numerical approach}.\textemdash
In our numerical study of Eq.~\eqref{eq:hamilt1}, we complement DMRG simulations with calculations based on NGS, $\ket{\Psi_\mathrm{NGS}}$, that can be written in the form \cite{shi17}
\begin{equation}
\ket{\Psi_\mathrm{NGS}} = U_S \ket{\Psi_\mathrm{GS}}
\end{equation}
where $U_S$ is a unitary operator and $\ket{\Psi_\mathrm{GS}}$ an arbitrary Gaussian state, both of which depend on a set of variational parameters \cite{SM}.
We derive and solve the equations of motion for these variational parameters to obtain the many-body ground state of $H_\mathrm{eff}$, see \cite{SM} for more details.
In order to treat the model in Eq.~\eqref{eq:hamilt1} with the NGS, we employ a Jordan-Wigner transformation and map $H_\mathrm{eff}$ onto a fermionic model via
\begin{equation}\label{eq:jordan-wigner-maintext}
    \sigma_i^z = 2c_i^\dagger c_i - 1, \ \sigma_i^+ = e^{i\pi\sum_{l<i}c_l^\dagger c_l}c_i^\dagger.
\end{equation}

Expressing the Hamiltonian \eqref{eq:hamilt1} in terms of fermionic operators by means of \eqref{eq:jordan-wigner-maintext} shows the similarity with the standard Holstein model, as studied in condensed matter physics.
In this analogy, spin-spin interactions translate to electron hopping and spin-phonon to electron-phonon interaction.
The differences between the standard Holstein model and our model \eqref{eq:hamilt1} are the following:
Firstly, one key difference originates from the long-range hopping terms $\propto P_{ij}/|i-j|^3 c_i^\dagger c_j$ (with the string operator $P_{ij}$, see \cite{SM} for more details) present in our effective fermionic model, which stems from the dipolar decay of interactions in Eq.~\eqref{eq:hamilt1}.
Secondly, in contrast to the genuine Holstein model which features a purely local coupling of electron and Einstein phonon, i.e. $g_{in} = \delta_{in}$ in Eq.~\eqref{eq:hInt}, the phonon described by Eq.~\eqref{eq:hamilt1} is dispersive and its bandwidth may be tuned by means of $\xi$.

While NGS excel at numerical efficiency and capture the essential physics well, DMRG yields higher numerical accuracy.
However the DMRG study of Eq.~\eqref{eq:hamilt1} faces several technical challenges.
Arguably two of the most relevant practical obstacles are associated with
(\textit{i}) not getting stuck in a local energy minimum during the algorithm, and
(\textit{ii}) avoiding truncation errors introduced by working with finite local phonon Hilbert spaces.
In our numerical treatment, we find that (\textit{i}) NGS can provide an excellent educated guess for the initial state fed into the DMRG algorithm, thus lowering the chances for getting stuck with a metastable solution.
Moreover, (\textit{ii}) the truncation error associated with finite local Hilbert spaces can be significantly lowered by employing a unitary displacement transformation on Eq.~\eqref{eq:hamilt1} (see \cite{SM}).
Note that more general approaches exist to tackle this issue and have been applied to problems with fermion-phonon coupling \cite{jeckelman98,zhang98,guo12,brockt15,stolpp20,koehler20}.

\textit{Phase diagram}.\textemdash
As the spin-spin couplings and spin-phonon interactions compete, the many-body ground state displays several distinct phases as a function of phonon parameter $\xi$ and spin-phonon coupling strength $F_z$.
Equipped with our numerical toolbox, we study the ground-state properties of $H_\mathrm{eff}$ and calculate several spin and phonon observables.
Especially, we introduce the CDW order parameter
\begin{equation}
    O_\mathrm{CDW} = \frac{1}{2N} \sum_{n=1}^N (-1)^n \left ( 1 + \langle \sigma_n^z \rangle \right ),
\end{equation}
and the four-point spin correlator
\begin{equation}
    O_\mathrm{SC} = \langle \sigma_i^+ \sigma_{i+1}^+ \sigma_{i+\delta}^- \sigma_{i+1+\delta}^- \rangle,
\end{equation}
with which we identify the superconducting ground state by calculating its decay as a function of $\delta$ for fixed $i$.
The order parameters that we compute with the NGS approach for the fermionic model are derived in the Supplemental Material \cite{SM}.

\begin{figure}[t!]
\includegraphics[width=\columnwidth]{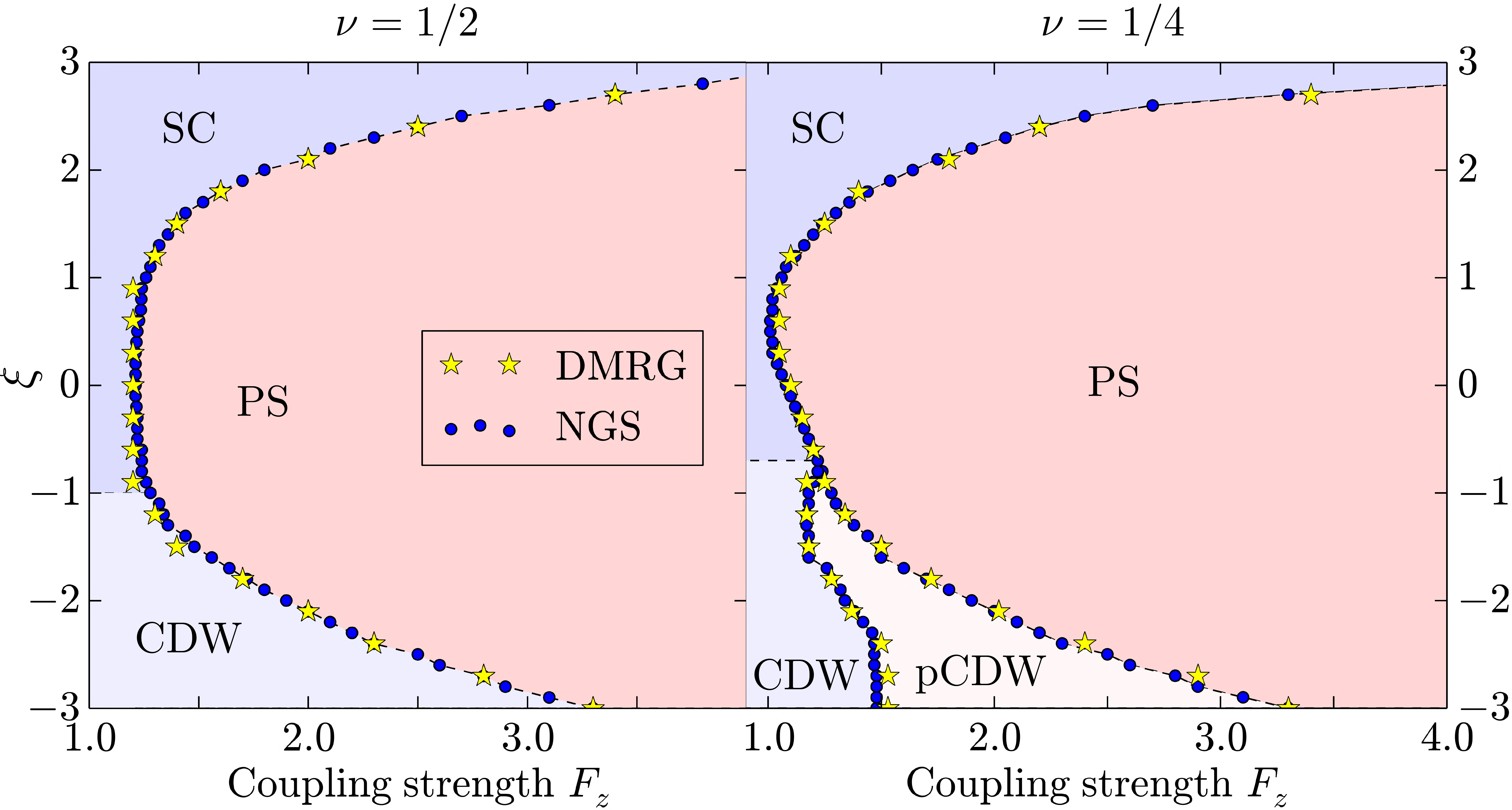}
\caption{\label{fig:phases}Phase diagram of spin-Holstein model.
$\xi \gtrsim 1$ ($\xi \lesssim -1$) corresponds to the soft (stiff) limit.
\textit{Left panel}: At filling factor $\nu = 1/2$, there exist three distinct phases at sufficiently large $F_z$, in a charge-density wave (CDW), a superconducting (SC) and a phase-separated (PS) regime.
\textit{Right panel}: At $\nu = 1/4$, there exists an additional pCDW phase (discussed in the main text).
Numerical parameters: $N = 48$, $\omega_z/J = 1$.
}
\end{figure}

We study the phase diagram for different filling factors $\nu = (\sum_i 1+\langle\sigma_i^z\rangle)/(2N)$.
In Fig.~\ref{fig:phases}, we show the result for $N = 48$ spins at $\nu = 1/2$ (left panel) and $\nu = 1/4$ (right panel) as a function of $F_z$ and $\xi$.
The phase boundaries obtained with both numerical methods quantitatively agree with each other.
Note that we focus here on the regime where $F_z\geq1$ since there exists only a trivial Luttinger-liquid phase at small couplings.

\begin{figure*}[t!]
\centering 
\includegraphics[width=0.72\textwidth]{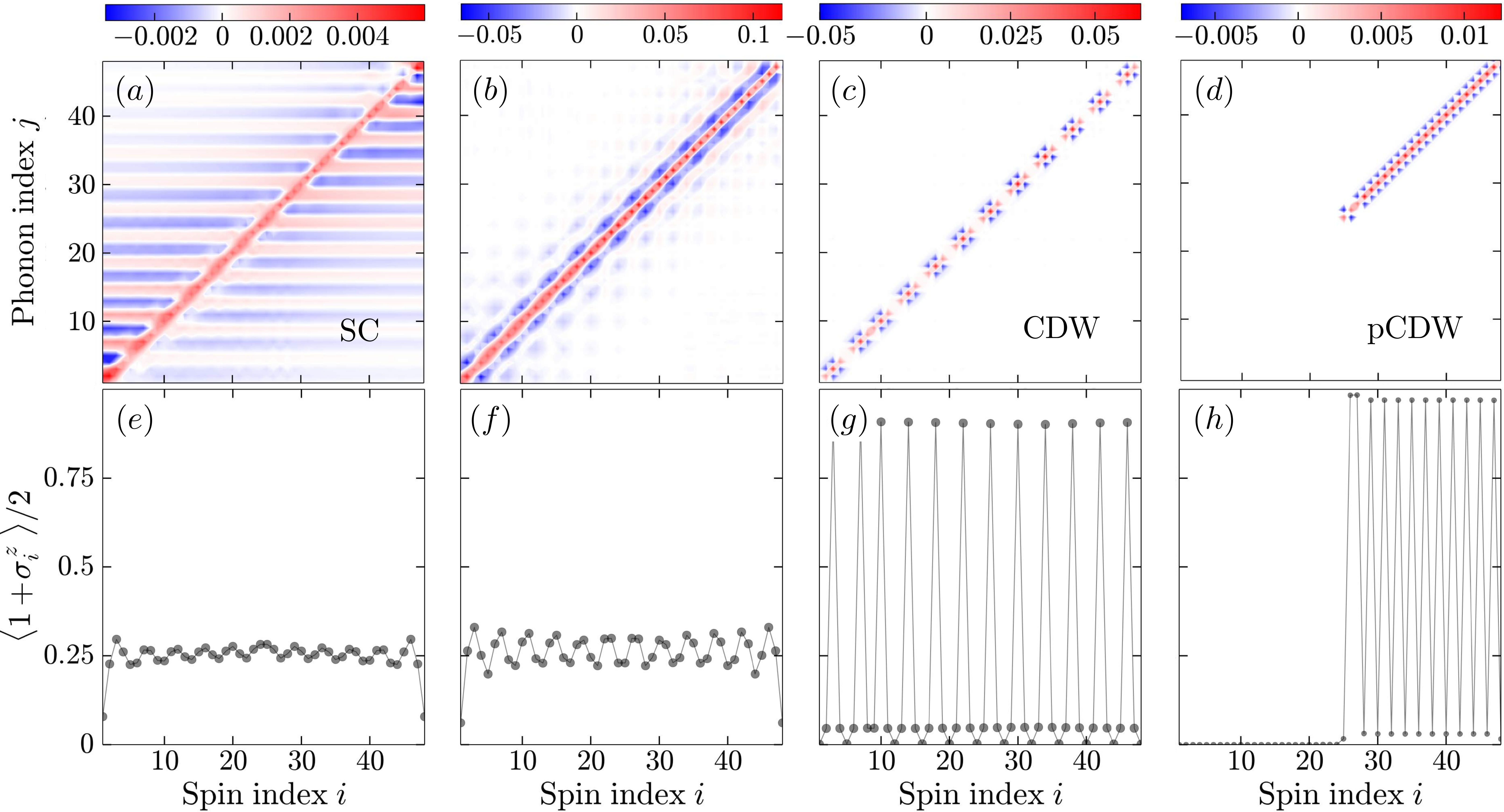}
\caption{\label{fig:spin-phonon}Spin phonon correlator $\Pi_{ij}$ and spin configuration for different ground states at quarter filling $\nu = 1/4$ as obtained with DMRG.
\textit{Upper panel}: Spin-phonon correlator $\Pi$ as defined in Eq.~\eqref{eq:spin-phonon-correlator}.
\textit{Lower panel}: Site-dependent spin expectation value $(1 + \langle \sigma_i^z \rangle)/2$.
The four columns correspond to particular choices for $F_z$ and $\xi$ (compare right panel of Fig.~\ref{fig:phases}).
(\textit{a}) and (\textit{e}): $F_z = 1.6$, $\xi = 2.1$ (SC regime).
(\textit{b}) and (\textit{f}): $F_z = 0.6$, $\xi = -2.1$ (precursor of CDW regime).
(\textit{c}) and (\textit{g}): $F_z = 1.2$, $\xi = -2.1$ (CDW regime).
(\textit{d}) and (\textit{h}): $F_z = 1.5$, $\xi = -2.1$ (pCDW regime).
Other numerical parameters: $N = 48$, $\omega_z/J = 1$.
}
\end{figure*}

At half filling ($\nu = 1/2$), and at sufficiently large spin-phonon coupling $F_z \gtrsim 1$, we find three distinct phases, that display charge-density wave order, quasi-long range superconducting order of \textit{p}-wave pairing, and phase separation into two regions, in which the spins are pointing either up or down, respectively:
(\textit{i}) In the stiff limit ($\xi \lesssim -1$), where the harmonic trapping potential dominates the Coulomb interaction, the phonons are more localized than in the soft limit.
As a result, the phonon fluctuations around the ions' equilibrium position are suppressed, and in the regime $\xi \lesssim -1$ we discover a CDW state as the preferred ground state at moderate $F_z$.
At half filling, the latter is characterized by an alternating spin configuration $\langle \sigma_n^z \rangle \propto (-1)^n$ ($n = 1, ..., N$) and a large order parameter $O_\mathrm{CDW} \sim 0.5$.
(\textit{ii}) In the soft limit ($\xi \gtrsim 1$),
where the virtual phonon fluctuations are large and responsible for inducing attractive pairing interactions,
we find a superconducting ground state that exhibits a slow power-law decay $O_\mathrm{SC} \sim \delta^{-\alpha}$, with $\alpha \approx 2$.
(\textit{iii})
There exists a competition between SC and CDW order, respectively, and phase separation.
At sufficiently large coupling $F_z$, the spin-Holstein model displays an instability towards phase separation into two regions with opposite polarization, both in the stiff and in the soft limit.

At quarter filling ($\nu = 1/4$), we map out a similar phase diagram, and find an additional phase in the stiff limit ($\xi \lesssim -1$), which we refer to as the pCDW phase as shorthand notation for a phase that displays both phase separation and CDW order, see Fig.~\ref{fig:phases}.
It is prevalent at intermediate coupling strength, and it is characterized by the coexistence of phase separation and an enhanced CDW order parameter,
with half of the spin chain being polarized and a staggered magnetization in the other half.
Representative results for the spin configurations of different phases at $\nu = 1/4$ are shown in Fig.~\ref{fig:spin-phonon}(\textit{e})-(\textit{h}).
%
In all cases, we find excellent agreement between the DMRG and NGS numerical results.

\textit{Spin-phonon correlations.}\textemdash
To study the correlation between spins and phonons, we calculate the observable
\begin{equation}\label{eq:spin-phonon-correlator}
    \Pi_{ij} = \langle \sigma_i^z r_j \rangle - \langle \sigma_i^z \rangle \langle r_j \rangle.
\end{equation}
In Figs.~\ref{fig:spin-phonon}(\textit{a})-(\textit{d}) we show the DMRG results at $\nu = 1/4$ which agree very well with the corresponding results obtained with the NGS.
In the superconducting regime, cf.~Fig.~\ref{fig:spin-phonon}(\textit{a}), the stripe pattern of $\Pi_{ij}$ demonstrates the presence of non-local spin-phonon correlations.
For a fixed spin index $i$, it displays oscillations with a period four near the center of the chain.
In contrast, the correlations decay quickly in the CDW and pCDW regimes and are symmetric about $i=j$.
In the stiff limit, at small couplings we find a precursor of the CDW state, where $\Pi_{ij}$ decays more slowly away from $i=j$ than deep in the CDW regime, compare Figs.~\ref{fig:spin-phonon}(\textit{b}) and (\textit{c}).
A representative result for $\Pi_{ij}$ for the pCDW ground state is shown in Fig.~\ref{fig:spin-phonon}(\textit{d}).
As expected, the spin-phonon correlations vanish in one half of the system, while in the other they feature oscillations with a period two along the diagonal $i=j$, as would be expected for a CDW state at half filling.
In the pCDW regime, the magnitude of the spin-phonon correlator is smaller than in the charge-density wave phase.
At even larger $F_z$ (phase separation), $\Pi_{ij}$ vanishes almost everywhere, except for small contributions close to the domain wall.

\textit{Phonon numbers}.\textemdash
To characterize the phonon excitations, we decompose the phonon excitation number into the density of coherent phonons, $n_c$, and quantum fluctuations of the phonon density, $n_s$:
\begin{equation}
n_{c}=\frac{1}{N}\sum_{k}\left\vert \left\langle a_{k}\right\rangle\right\vert^{2}, \ \ n_{s}= \frac{1}{N}\sum_{k}\left\langle a_{k}^{\dagger }a_{k}\right\rangle
-n_{c}.
\end{equation}
For Fig.~\ref{fig:spin-phonon}, the average phonon numbers are (a) $n_c=0.64$ and $n_s=0.51$, (b) $n_c$=0.093 and $n_s$=0.52, (c) $n_c$=0.65 and $n_s$=0.75 (d) $n_c$=2.06 and $n_s$=0.5.
When the system is in the SC phase, the virtual phonon fluctuations induce attractive interactions necessary for pairing, as familiar from BCS theory of superconductivity.
In contrast, as the system enters the pCDW phase, the coherent phonon displacement becomes dominant.
In the normal phase, the displacement is very small.

\textit{Experimental considerations}.\textemdash
Trapped-ion experiments benefit from well-developed readout techniques.
Ions can be excited from one spin state to another with single-site resolution, and subsequent fluorescence imaging allows the extraction of local expectation values $\langle \sigma_i^z \rangle$.
Repeated measurements at different sites enable access to spin-spin correlation functions like $\langle \sigma_i^z \sigma_j^z \rangle$ and $O_\mathrm{SC}$.
Spin-phonon correlations may be probed with only spin measurements and additional lasers that locally couple spins and phonons.
All observables of our numerical study may thus be probed experimentally.
Many recent experiments have demonstrated that trapped-ion quantum simulations of spin models are feasible, with system sizes comparable to those considered here \cite{zhang17,monroe21}.

While we have only shown numerical results for a system with $N=48$ ions, we also study how the phase boundary in Fig.~\ref{fig:phases} shifts in the $(F_z, \xi)$ plane with respect to the system size $N$ using NGS.
Deep in the stiff limit ($\xi = -3$), we find that there is no noticeable influence of the system size on the phase boundaries both at half and quarter filling factors.
However in the soft limit, we find that the phase boundary moves to smaller (larger) $F_z$ as $N$ is increased (decreased).
For example, at $\nu = 1/2$ and $\xi = 3$, we find the SC-to-PS transition near $F_z = 2.6$ for $N = 96$ and $F_z = 4.5$ for $N = 48$,
while for $N = 24$, the phase boundary disappears, i.e., we do not find any critical point numerically for $F_z \leq 16$.
For larger systems and in the soft limit, smaller coupling strengths are thus sufficient to induce phase separation.
A scaling analysis of the NGS results obtained for systems with up to $N = 400$ shows that the SC phase survives in the thermodynamic limit.
For example, at $\xi = 3$ the SC-to-PS boundary moves to $F_z \approx 1$ for $N \rightarrow \infty$.

We perform finite-temperature calculations using the NGS to confirm that the predicted phases survive at $T > 0$ and may actually be observed in state-of-the-art experiments.
At temperatures up to $T \sim J / k_\mathrm{B}$, we find that the $T = 0$ ground states are robust and the phase diagrams in Fig.~\ref{fig:phases} change only insignificantly.
For ${}^{40}\mathrm{Ca}^+$ ions at an effective temperature $T = 1\mu$K and with our choice $\omega_z/J = 1$, this corresponds to trap distances $d_0 \approx 5\mu$m deep in the soft limit ($\xi = 3$) and larger separations in the stiff limit.
This shows that our results are consistent with the parameters of typical trapped-ion setups.

\textit{Conclusions.}\textemdash
To conclude, we have studied a generalized Holstein model that can be implemented in state-of-the-art trapped-ion experiments.
In our numerical study, we have demonstrated that it can be useful to choose a hybrid approach in which calculations based on non-Gaussian variational ansatz states and density-matrix renormalization group complement each other.
This allowed us to map out the phase diagram of the trapped-ion spin system, which is governed solely by tunable laser and ion trap parameters.
While we have concentrated on $\nu = 1/2$ and $\nu = 1/4$, other filling factors may be explored in future work, and could give rise to an even richer hierarchy of phases in the stiff limit.
As a future prospect, also more exotic models could be investigated that include higher-order interactions between phonons and spins.
While they would be harder to tackle with classical methods, in a trapped-ion quantum simulator, they may be implemented by driving higher-order sidebands with a laser.
That way, the quantum simulator may possibly be operated in two regimes, one which is also accessible with classical calculations, and another that may go beyond what's achievable with state-of-the-art numerics.
A straightforward extension of our work is the consideration of well-established Paul trap setups instead of microtrap arrays, in which the ions are not perfectly equidistantly spaced.

\begin{acknowledgments}
\textit{Acknowledgements}.\textemdash
J.K. and J.I.C. acknowledge support from the Deutsche Forschungsgemeinschaft (DFG, German Research Foundation) under Germany’s Excellence Strategy – EXC-2111 – 390814868.
T.S. is supported by the NSFC (Grants No. 11974363).
E.D. acknowledges support from the ARO grant number W911NF-20-1-0163, the AFOSR-MURI award FA95501610323, and the NSF EAGER-QAC-QSA award number 2222-206-2014111.
J.K. thanks Miles Stoudenmire for useful discussions.

J.K. and T.S. contributed equally to this work.
\end{acknowledgments}

\clearpage

\widetext

\begin{center}
\textbf{\large Supplemental Materials: Spin-Holstein models in trapped-ion systems}
\end{center}

This Supplemental Material is structured as follows.
Sec.~\ref{sm:effective} summarizes the derivation of the effective Hamiltonian used in the main text.
In Sec.~\ref{sm:displ} we discuss the displacement transformation employed for the DMRG simulations.
Numerical convergence of these simulations is discussed in Sec.~\ref{sm:convergence}. 
We discuss the fermionic model and non-Gaussian state ansatz in Sec.~\ref{sm:ngs}.
In Sec.~\ref{sm:other-observables}, we complement the results from the main text with additional numerical data on the structure factor $S(q)$, phonon observables and the domain wall in the phase-separated regime.

\setcounter{equation}{0}
\setcounter{figure}{0}
\setcounter{table}{0}
\setcounter{page}{1}
\setcounter{section}{0}
\makeatletter
\renewcommand{\theequation}{S\arabic{equation}}
\renewcommand{\thefigure}{S\arabic{figure}}
\renewcommand{\bibnumfmt}[1]{[S#1]}
\renewcommand{\citenumfont}[1]{S#1}
\renewcommand{\thesection}{S\arabic{section}}%
\setcounter{secnumdepth}{3}

\section{Derivation of effective Hamiltonian $H_\mathrm{eff}$ \label{sm:effective}}

In the following we summarize the derivation of the spin-Holstein model as studied in the main text.
We discuss the validity of the underlying approximations.

\subsection{Phonons}

The vibrations of ions in a microtrap array can be described by ($\hbar = 1$)
\begin{equation}\label{eq:phonons}
    H_\mathrm{ph} = \sum_{\substack{i=1,\\ \alpha=x,y,z}}^N \frac{(p_i^\alpha)^2}{2m} + \frac{m}{2} \sum_{i,j,\alpha} \mathcal{K}_{ij}^\alpha r_i^\alpha r_j^\alpha,
\end{equation}
where $p_i^\alpha$ denotes the momentum and $r_i^\alpha = 1/\sqrt{2m\omega_\alpha} (b_{i,\alpha} + b_{i,\alpha}^\dagger)$ is the displacement of the $i$th ion from its equilibrium position in the $\alpha$ (= $x, y$ or $z$) direction, with the trap frequencies $\omega_\alpha$ and local phonon ladder operators $b_{i,\alpha}^{(\dagger)}$.
$\mathcal{K}^\alpha$ denotes the elasticity matrix of the ion chain in the $\alpha$ direction, and its eigenvectors describe the chain's normal modes \cite{porras04}.

\subsection{Spin-phonon interaction}

In the setup we consider (see main text), a laser beam configuration is chosen to host three standing waves along the $\hat x$, $\hat y$ and $\hat z$ axes.
Light that is off-resonant with chosen hyperfine-state transitions of the ions can be harnessed to introduce a coupling between the motional and spin degrees of freedom of the ions \cite{porras04,deng05}.
Within the rotating-wave approximation, we obtain an effective spin-phonon coupling of the form
\begin{equation}\label{eq:interaction-hamilt}
H_\mathrm{int} = \sum_{\substack{i=1,\\ \alpha=x,y,z}}^N \frac{\Omega_\alpha^2}{2\Delta_\alpha} \cos^2 \left ( k_\alpha r_i^\alpha + \phi_\alpha \right ) \left ( 1 + \sigma_i^\alpha \right ),
\end{equation}
where $\Omega_\alpha$ denotes the Rabi frequency, $\Delta_\alpha$ is the qubit-laser detuning, $k_\alpha$ the wavenumber of the standing light wave, $\phi_\alpha$ the relative phase of counterpropagating lasers and $\sigma_i^\alpha$ denotes the Pauli matrix associated with the internal spin states $\ket{\uparrow}$ and $\ket{\downarrow}$ at site $i$ and direction $\alpha$.
In the Lamb-Dicke regime, characterized by a small parameter $\eta_\alpha = k_\alpha/\sqrt{2m\omega_\alpha} \ll 1$, Eq.~\eqref{eq:interaction-hamilt} can be linearized around the ions' equilibrium positions, so that the interaction Hamiltonian takes the simplified form $H_\mathrm{int} = - F_\alpha \sum_{i,\alpha} r_i^\alpha (1 + \sigma_i^\alpha)$, with a coupling strength $F_\alpha \sim \Omega_\alpha^2 k_\alpha / (2\Delta_\alpha)$ that can be controlled by laser parameters.

\subsection{Effective spin-Holstein model}

For completeness, we sketch here the derivation of our effective model based on the phonon part \eqref{eq:phonons} and light-induced spin-phonon interaction \eqref{eq:interaction-hamilt}.
We refer to the existing trapped-ion literature for more details, cf.~Refs.~\cite{porras04,deng05}.
We start from Eqs.~\eqref{eq:phonons} and \eqref{eq:interaction-hamilt} in the main text and an additional external magnetic-field term,
\begin{equation}\label{eq:model-SM}
    H = \sum_{i=1}^N \sum_{\alpha=x,y,z} \frac{(p_i^\alpha)^2}{2m} + \frac{m}{2} \sum_{i,j=1}^N\sum_{\alpha=x,y,z} \mathcal{K}_{ij}^\alpha r_i^\alpha r_j^\alpha +
    \sum_{i=1}^N\sum_{\alpha=x,y,z} \frac{\Omega_\alpha^2}{2\Delta_\alpha} \cos^2 \left ( k_\alpha r_i^\alpha \right ) \left ( 1 + \sigma_i^\alpha \right ) + \sum_{i=1}^N \sum_{\alpha=x,y,z} B_\alpha \sigma_i^\alpha,
\end{equation}
where the elasticity matrix is given by \cite{porras04}
\begin{equation}
\mathcal{K}_{ij}^\alpha =
\omega_z^2 \times \begin{cases} 1 + c_\alpha \underset{{k\neq i}}{\sum} \frac{\beta_\alpha}{|i-k|^3}, & i = j,\\
-c_\alpha \frac{\beta_\alpha}{|i-j|^3}, & i \neq j,
\end{cases}
\end{equation}
with $c_{x,y} = 1$ and $c_z = - 2$.
Now we apply a unitary transformation $U_\mathrm{pol}$ to Eq.~\eqref{eq:model-SM},
\begin{equation}
    U_\mathrm{pol} = \exp \left ( \sum_{i=1}^N \sum_{n=1}^N \sum_{\alpha = x, y} \eta_{i,n}^\alpha (1 + \sigma_i^\alpha) ( a_{n,\alpha}^\dagger - a_{n,\alpha} ) \right ),
\end{equation}
where
\begin{equation}
\eta_{i,n}^\alpha = \frac{F_\alpha}{\Omega_{n,\alpha}\sqrt{2m\Omega_{n,\alpha}}} \mathcal{M}_{i,n}^\alpha, \ \ \ \sum_{i,j=1}^N \mathcal{M}_{i,n}^\alpha \mathcal{K}_{ij}^\alpha \mathcal{M}_{j,m}^\alpha = \Omega_{n,\alpha}^2 \delta_{mn}.
\end{equation}
Denoting with $H_\mathrm{eff} = U_\mathrm{pol} H U_\mathrm{pol}^\dagger$ our effective Hamiltonian, to first order in $\eta_{i,n}^\alpha$ we eliminate the transverse ($\alpha = x, y$) phonons and interaction terms from the description, and introduce effective spin-spin interaction terms.
The Hamiltonian takes the form
\begin{equation}\label{eq:deriv-2}
    H_\mathrm{eff} = \sum_{n} \Omega_n a_n^\dagger a_n - F_z \sum_{i=1}^N \sum_{n=1}^N g_{in} (a_{n,z} + a_{n,z}^\dagger )(1 + \sigma_i^z) + \sum_{i\neq j} \sum_{\alpha=x,y} J_{ij}^\alpha \sigma_i^\alpha \sigma_j^\alpha + \sum_{i=1}^N \sum_{\alpha=x,y,z} \left ( B_\alpha - \frac{F_\alpha^2}{m \omega_\alpha^2} \right ) \sigma_i^\alpha,
\end{equation}
where $g_{in} = \mathcal{M}_{in}/\sqrt{2m\Omega_n}$.
Note that within the Lamb-Dicke approximation we have neglected terms that are of second and higher order in $\eta^\alpha_{i,n}$ \cite{porras04}.

The last term in Eq.~\eqref{eq:deriv-2} shows why we introduced external magnetic fields in Eq.~\eqref{eq:model-SM}.
The global force term stemming from the transformation can be canceled by appropriately choosing $B_\alpha$ along all directions.
Note that in the main text we focus on the case where $B_\alpha - F_\alpha^2/(m\omega_\alpha^2) = 0$ along all three directions.

\begin{figure}[t!]
\includegraphics[width=0.4\columnwidth]{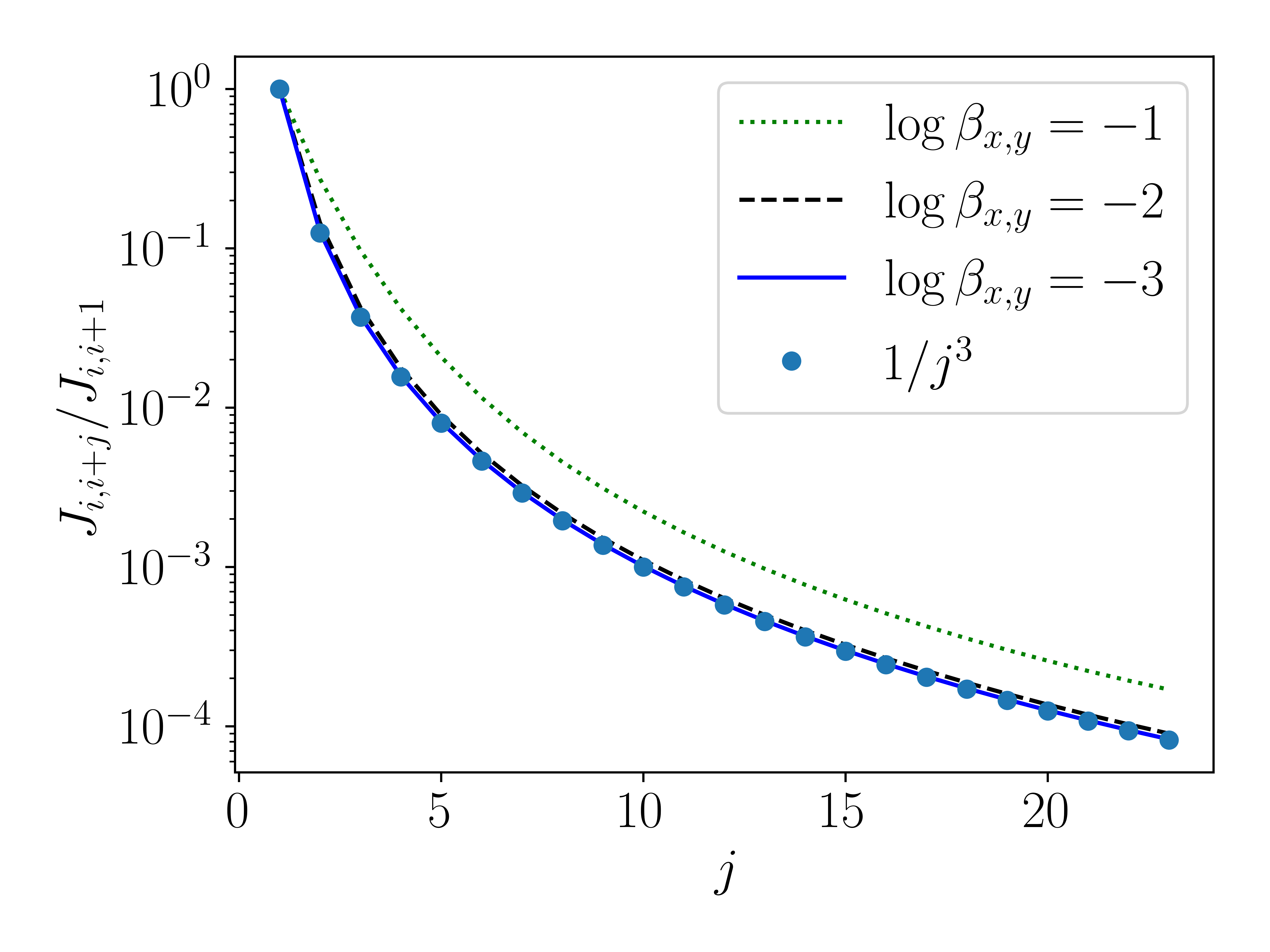}
\caption{\label{fig:Jij}Scaling $J_{ij} \sim 1/|i-j|^3$ in the stiff limit.
Result shown for chain of length $N=48$ and distance from central spin at site $i=24$.
}
\end{figure}

In order to obtain the effective model (1) from the main text, we use that $J_{ij} \equiv J_{ij}^x = J_{ij}^y \sim 1/|i-j|^3$.
In general, the interaction can be derived from the elasticity matrix,
\begin{equation}\label{eq:SM-Jij}
    J_{ij}^\alpha = - \frac{F_\alpha^2}{m} \left [ \left ( \mathcal{K}^\alpha \right )^{-1} \right ]_{ij}.
\end{equation}
Since we have assumed large transverse trap frequencies to adiabatically eliminate transverse phonons, the transverse traps are operated in the stiff limit.
In this case, the dipolar scaling $J_{ij} = 1/|i-j|^3$ follows directly from \eqref{eq:SM-Jij}.
Fig.~\ref{fig:Jij} demonstrates the scaling for decreasing $\beta_{x,y}$ in the stiff limit.

\newpage
\subsection{Spin-phonon coupling matrix}

\begin{figure}[t!]
\includegraphics[width=0.65\columnwidth]{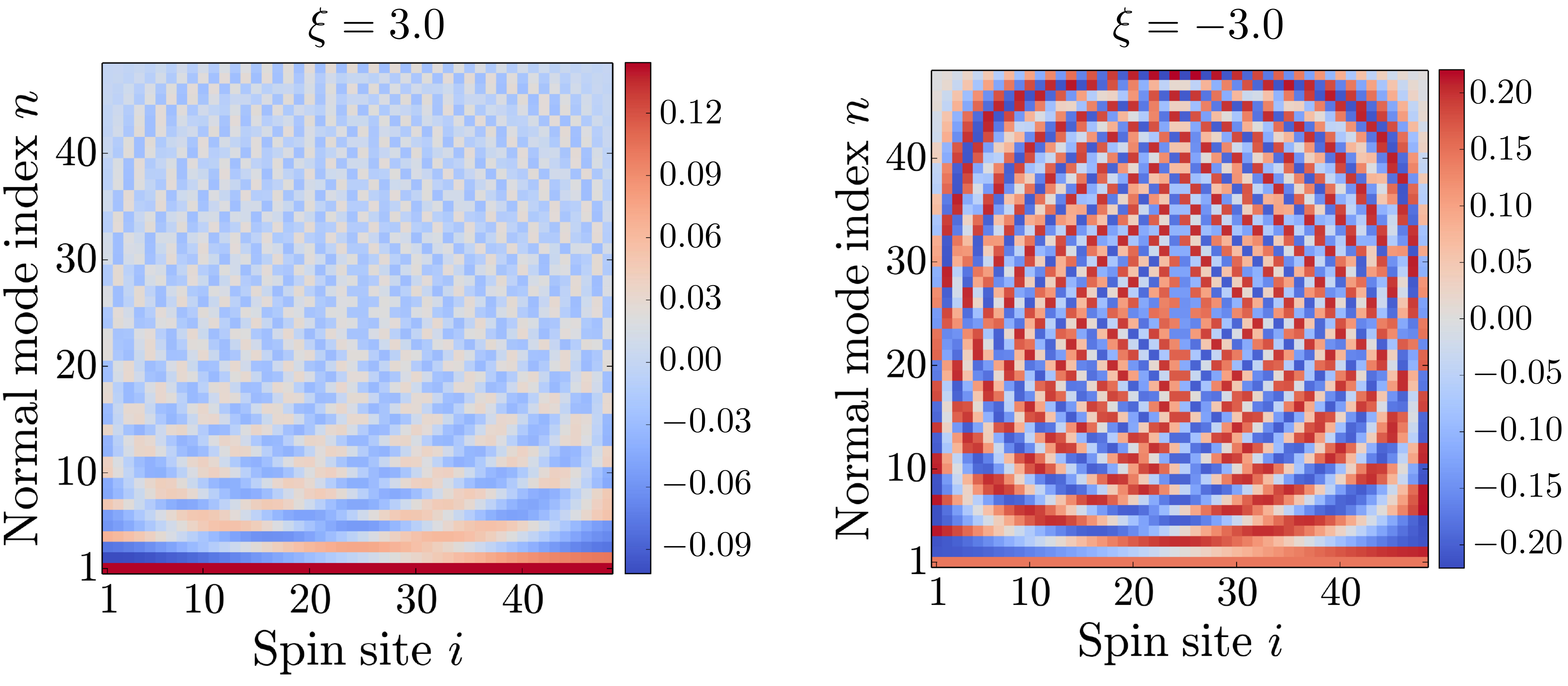}
\caption{\label{fig:gin}Coupling strength $g_{in}$ for $\xi = 3.0$ in the soft limit (left) and $\xi = -3.0$ in the stiff limit (right).
Result shown for chain of length $N=48$.
}
\end{figure}

The exact form of the non-local coupling between spins and phonons depends on the stiffness of the ion traps, as described by $\xi$.
While the form of $g_{in}$ for $\xi = 0$ is shown in the main text, we complement that result by showing $g_{in}$ in the stiff and soft limits in Fig.~\ref{fig:gin}.

\section{Displacement transformation \label{sm:displ}}
In our numerical simulations, it is often more convenient to remove the phononic displacement term $\sim F_z \sum_{i,n} g_{in} (a_n + a_n^\dagger)$ from the description at the cost of a spin-dependent shift.
It renders the DMRG calculations more efficient, especially at strong coupling $F_z$.
To this aim, we introduce the displaced phonon operators
\begin{equation}
\hat \alpha_n := \hat a_n - F_z \sum_i \frac{\mathcal{M}_{in}}{\sqrt{2m}\Omega_n^{3/2}} \equiv \hat a_n - c_n.
\end{equation}
As the displacement transformation only affects the phonons, we denote the spin Hamiltonian by $H_s = \sum_{i,j,\alpha} J_{ij} \sigma_i^\alpha \sigma_j^\alpha$
and rewrite Eq.~\eqref{eq:deriv-2} as
\begin{eqnarray}
H_\mathrm{eff} & = & \sum_{n} \Omega_n a_n^\dagger a_n - F_z \sum_{i,n} \frac{\mathcal{M}_{in}}{\sqrt{2m\Omega_n}} (a_n + a_n^\dagger) (1 + \sigma_i^z) + H_s \nonumber \\
& = & \sum_{n} \Omega_n \alpha_n^\dagger \alpha_n + \sum_n \Omega_n c_n (\alpha_n + \alpha_n^\dagger) + \sum_n \Omega_n c_n^2 \nonumber \\
& & - F_z \sum_{i,n} \frac{\mathcal{M}_{in}}{\sqrt{2m\Omega_n}} (\alpha_n + \alpha_n^\dagger) - F_z \sum_{i,n} \frac{\mathcal{M}_{in}}{\sqrt{2m\Omega_n}} (\alpha_n + \alpha_n^\dagger) \sigma_i^z - 2 F_z \sum_{i,n} \frac{\mathcal{M}_{in}}{\sqrt{2m\Omega_n}}c_n - 2 F_z \sum_{i,n} \frac{\mathcal{M}_{in}}{\sqrt{2m\Omega_n}}c_n \sigma_i^z + H_s \nonumber \\
& = & \sum_{n} \Omega_n \alpha_n^\dagger \alpha_n - F_z \sum_{i,n} g_{in} (\alpha_n + \alpha_n^\dagger) \sigma_i^z + H_s + H_\mathrm{R},
\end{eqnarray}
with the residual phonon-independent contribution
\begin{equation}
    H_\mathrm{R} = - F_z^2 \sum_{n} \frac{1}{2m \Omega_n^2} \left ( \sum_i \mathcal{M}_{in} \right )^2 - 2 F_z^2 \sum_{i,n}\frac{\mathcal{M}_{in}}{2m\Omega_n^2}\sigma_i^z \left ( \sum_j \mathcal{M}_{jn} \right ).
\end{equation}

\section{Convergence analysis \label{sm:convergence}}

\begin{figure}[t!]
\includegraphics[width=0.7\columnwidth]{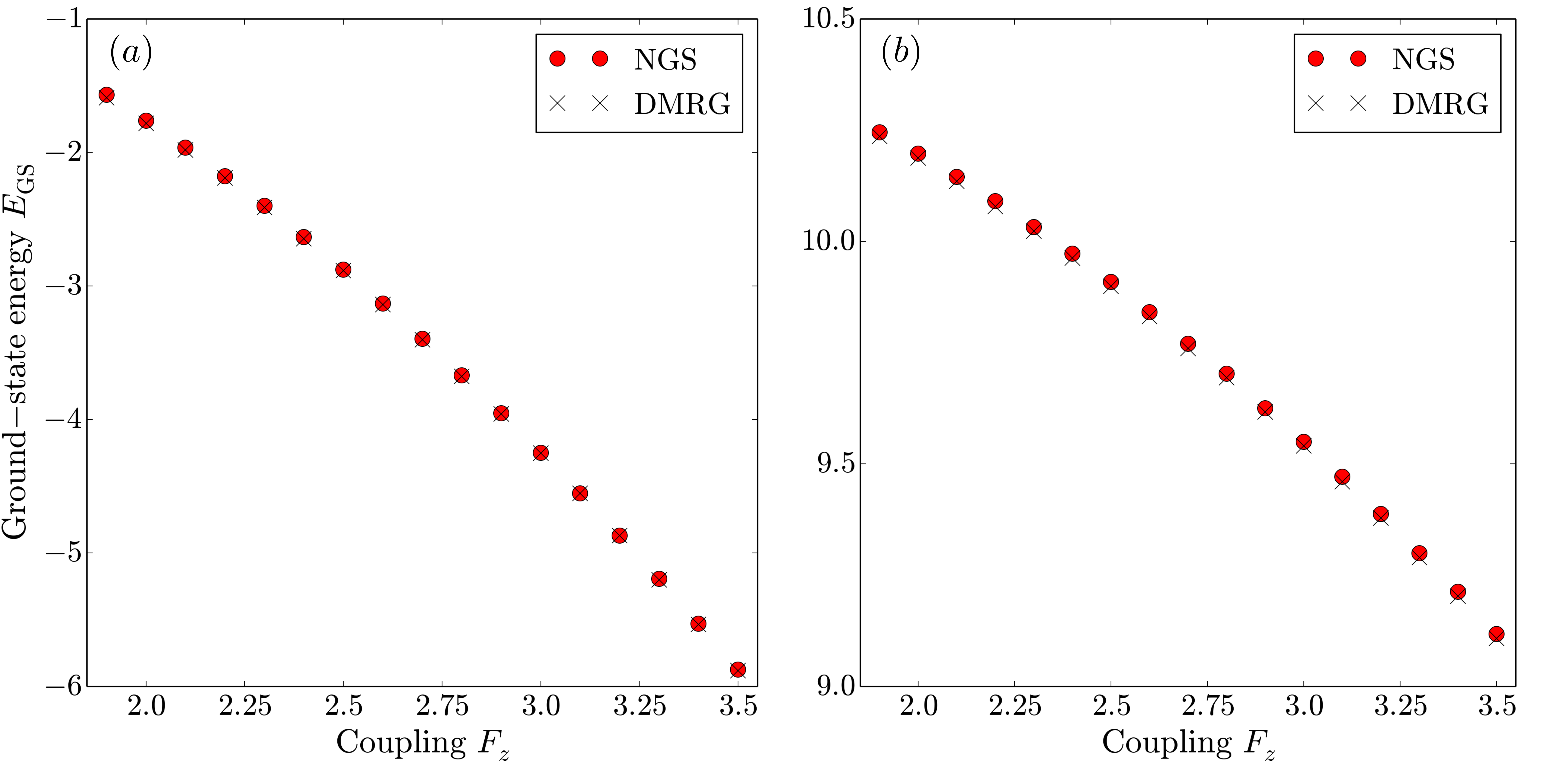}
\caption{\label{fig:energies_SM}Ground-state energies obtained with DMRG and NGS methods.
(\textit{a}) $\beta = -3.0$ (stiff) and (\textit{b}) $\beta = 3.0$ (soft).
Other numerical parameters: $N = 48$, $\omega_z / J = 1$.
}
\end{figure}

We benchmark our numerical calculations against each other and compare the ground state energies obtained with DMRG and the NGS method outlined in Sec.~\ref{sm:ngs}.
Most importantly, we make sure that the ground state energies are close to each other, see Fig.~\ref{fig:energies_SM}.
Typically the energies obtained with DMRG are slightly lower.
However, the runtime of the simulation is decreased significantly if we first perform the NGS calculation and then feed a good initial seed into the DMRG simulation.

In the derivation of the phase diagram (see Fig.~2 in the main text) we explore the ground states in $(F_z, \beta)$ space using an adaptive grid with a higher resolution closer to the phase boundary.
Slight deviations between the NGS and DMRG results can partially be explained by the fact that the maximal resolution used within the NGS calculations can be as small as $\Delta F_z \sim 10^{-4}$, while in the DMRG simulations we limit ourselves to $\Delta F_z \sim 10^{-2}$.
Similarly, with DMRG we discretize the possible values for the stiffness parameter and choose a resolution $\Delta \nu_\beta = 0.3$.
In our numerical calculations based on non-Gaussian ansatz states we choose $\Delta \nu_\beta = 0.1$ instead, as they are less costly.

\section{Non Gaussian states and equations of motion \label{sm:ngs}}

In this section, we study the ground state and thermal properties of a 1D
array of ions with the lattice spacing $d$ and mass $m$, which is described
by the 1D spin-Holstein model%
\begin{eqnarray}
H &=&\sum_{i\neq j,\alpha =x,y}J_{ij}\sigma _{i}^{\alpha }\sigma
_{j}^{\alpha }+\frac{B}{2}\sum_{i}\sigma _{i}^{z} \\
&&+\sum_{i}\frac{p_{i}^{2}}{2m}+\frac{m}{2}\sum_{ij}\mathcal{K}%
_{ij}r_{i}r_{j}-\sum_{i}F_{z}r_{i}(1+\sigma _{i}^{z}).  \notag
\end{eqnarray}%
The long-range interaction $J_{ij}=J_{\mathrm{dd}}/\left\vert i-j\right\vert
^{3}$ between ions at sites $i$ and $j$ is induced by the transverse phonon
modes, where the exchange interaction strength $J_{\mathrm{dd}%
}=F_{0}^{2}e^{2}/(2m^{2}\omega _{0}^{4}d^{3})$ is determined by the ion
charge $e$, the frequency $\omega _{0}$ of the microtrap and the force $%
F_{0} $ generated by the laser along the transverse directions $\alpha =x,y$%
. The longitudinal mode is described by%
\begin{equation}
\mathcal{K}_{ij}=(\omega _{z}^{2}+\sum_{l\neq i}\frac{\omega _{\mathrm{dd}%
}^{2}}{\left\vert i-l\right\vert ^{3}})\delta _{ij}-\frac{\omega _{\mathrm{dd%
}}^{2}}{\left\vert i-j\right\vert ^{3}}(1-\delta _{ij}),
\end{equation}%
where $\omega _{\mathrm{dd}}=\sqrt{2e^{2}/(md^{3})}$. The laser along the
longitudinal direction induces the local Holstein interaction between the
internal state and the longitudinal mode\ with strength $F_{z}$.

Via the Jordan-Wigner transformation%
\begin{eqnarray}
\sigma _{i}^{z} &=&2c_{i}^{\dagger }c_{i}-1,  \notag \\
\sigma _{i}^{+} &=&e^{i\pi \sum_{l<i}c_{l}^{\dagger }c_{l}}c_{i}^{\dagger },
\end{eqnarray}%
we rewrite the Hamiltonian%
\begin{eqnarray}
H &=&\sum_{i\neq j}\frac{1}{\left\vert i-j\right\vert ^{3}}%
P_{ij}c_{i}^{\dagger }c_{j}+B\sum_{i}c_{i}^{\dagger }c_{i} \\
&&+\frac{1}{4}\sum_{i}\bar{p}_{i}^{2}+\frac{1}{4}\sum_{ij}\mathcal{K}_{ij}%
\bar{r}_{i}\bar{r}_{j}-2\sum_{i}\bar{F}_{z}\bar{r}_{i}c_{i}^{\dagger }c_{i},
\notag
\end{eqnarray}%
where $t_{0}=4J_{\mathrm{dd}}$ is chosen as the unit, $\bar{F}_{z}=F_{z}/%
\sqrt{2m}$, and%
\begin{equation}
\bar{r}_{i}=\sqrt{2m}r_{i},\bar{p}_{i}=\sqrt{\frac{2}{m}}p_{i}
\end{equation}%
satisfy the canonical commutation relation $[\bar{r}_{i},\bar{p}_{i}]=2i$.
The operator $P_{ij}=e^{i\pi \sum_{l\in \mathcal{S}_{ij}}c_{l}^{\dagger
}c_{l}}$ is defined on the string $\mathcal{S}_{ij}$ connecting sites $i$
and $j$ (without points $i$ and $j$).

We employ the variational ansatz%
\begin{eqnarray}\label{eq:ngs-ansatz-appendix}
\left\vert \Psi _{\mathrm{NGS}}\right\rangle &=&U_{S}\left\vert \Psi _{%
\mathrm{GS}}\right\rangle , \\
\rho _{\mathrm{NGS}} &=&U_{S}\rho _{\mathrm{GS}}U_{S}^{\dagger },
\end{eqnarray}%
combining the generalized Lang-Firsov transformation%
\begin{equation}\label{eq:unitary-appendix}
U_{S}=e^{i\sum_{nl}\lambda _{ln}\bar{p}_{l}c_{n}^{\dagger }c_{n}}
\end{equation}%
and the Gaussian states%
\begin{eqnarray}
\left\vert \Psi _{\mathrm{GS}}\right\rangle &=&e^{-\frac{1}{2}R^{T}\sigma
^{y}\Delta _{R}}e^{-i\frac{1}{4}R^{T}\xi _{b}R}e^{i\frac{1}{2}C^{\dagger
}\xi _{f}C}\left\vert 0\right\rangle , \\
\rho _{\mathrm{GS}} &=&e^{-\frac{1}{2}R^{T}\sigma ^{y}\Delta _{R}}\frac{1}{%
Z_{b}}e^{-\Omega _{b}}e^{\frac{1}{2}R^{T}\sigma ^{y}\Delta _{R}}\frac{1}{%
Z_{f}}e^{-\Omega _{f}}
\end{eqnarray}%
to study the ground state and the thermal state, where the partition
functions $Z_{b,f}=tre^{-\Omega _{b,f}}$. The Gaussian state $\left\vert
\Psi _{\mathrm{GS}}\right\rangle $ and $\rho _{\mathrm{GS}}$ is completely
characterized by the covariance matrices $\Gamma _{f}=\left\langle
CC^{\dagger }\right\rangle $ (equivalently, $\Gamma _{m}=i(W_{f}\Gamma
_{f}W_{f}^{\dagger }-1)$ in the Majorana basis $A=W_{f}C$) and $\Gamma
_{b}=\left\langle \{R,R^{T}\}\right\rangle /2$ defined in the basis $%
C=(c_{i},c_{i}^{\dagger })^{T}$ and $R=(\bar{r}_{i},\bar{p}_{i})^{T}$, where 
$W_{f}=\left( 
\begin{array}{cc}
1 & 1 \\ 
-i & i%
\end{array}%
\right) $.

For the imaginary time evolution,\ the projections of%
\begin{eqnarray}
\partial _{\tau }\left\vert \Psi _{\mathrm{NGS}}\right\rangle &=&-\mathbf{P}%
(H-\left\langle H\right\rangle )\left\vert \Psi _{\mathrm{NGS}}\right\rangle
, \\
\partial _{\tau }\left\vert \Psi _{\rho }\right\rangle &=&-\mathbf{P}%
(F-f)\left\vert \Psi _{\rho }\right\rangle ,
\end{eqnarray}%
on the tangential space result in the EOM of variational parameters $\lambda
_{ln}$, $\Delta _{R}$, and $\Gamma _{f,b}$, which in the limit $\tau
\rightarrow \infty $ give rise to the ground state and thermal state,
respectively. Here, $\left\vert \Psi _{\rho }\right\rangle =U_{S}\otimes
I\left\vert \rho _{\mathrm{GS}}\right\rangle $ is determined by the
purification $\left\vert \rho _{\mathrm{GS}}\right\rangle =\sqrt{\rho _{%
\mathrm{GS}}}\otimes I\left\vert \phi ^{+}\right\rangle $ of the Gaussian
state $\rho _{\mathrm{GS}}$ via the maximal entangled state $\left\vert \phi
^{+}\right\rangle $ between the physical space and the fiducial space, and
the free energy $f$ is the average value of the free energy operator $%
F=H+T\ln \rho _{\mathrm{GS}}$.

The flow equations of variational parameters are%
\begin{equation}
\partial _{\tau }\Delta _{x}=-(\Gamma _{b}^{-1})_{\mathrm{phys},x}^{-1}(%
\mathcal{K}\Delta _{x}-2\sum_{n}G_{ln}\left\langle c_{n}^{\dagger
}c_{n}\right\rangle )+2\sum_{n}\partial _{\tau }\lambda _{ln}\left\langle
c_{n}^{\dagger }c_{n}\right\rangle ,
\end{equation}%
\begin{eqnarray}
\partial _{\tau }\Gamma _{b} &=&\sigma ^{y}\Omega \sigma ^{y}-\Gamma
_{b}\Omega \Gamma _{b}, \\
\partial _{\tau }\Gamma _{f} &=&\{\mathcal{F}_{f},\Gamma _{f}\}-2\Gamma _{f}%
\mathcal{F}_{f}\Gamma _{f},
\end{eqnarray}%
and%
\begin{equation}
\partial _{\tau }\lambda _{ln}=\sum_{n^{\prime }\neq m}w_{l,n^{\prime }m}%
\frac{e^{-\frac{1}{2}w_{n^{\prime }m}^{T}\Gamma _{p}w_{n^{\prime }m}}}{%
\left\vert n^{\prime }-m\right\vert ^{3}}\left\langle P_{n^{\prime
}m}c_{n^{\prime }}^{\dagger }c_{m}\right\rangle D_{n^{\prime
}n}^{-1}-(\Gamma _{b,p}^{-1}G)_{ln}
\end{equation}%
with $D_{nm}=\left\langle c_{n}c_{m}^{\dagger }\right\rangle \left\langle
c_{n}^{\dagger }c_{m}\right\rangle +\left\langle c_{n}^{\dagger
}c_{m}^{\dagger }\right\rangle \left\langle c_{m}c_{n}\right\rangle $ and $%
G_{ln}=2F_{z}\delta _{ln}+(\mathcal{K}\lambda )_{ln}$, where for phonons the
effective mean-field matrix%
\begin{equation}
\Omega =\left( 
\begin{array}{cc}
\mathcal{K} & 0 \\ 
0 & \Omega _{p}%
\end{array}%
\right) -T\sigma ^{y}\ln \frac{\Gamma _{b}\sigma ^{y}+1}{\Gamma _{b}\sigma
^{y}-1}
\end{equation}%
is determined by%
\begin{equation}
\Omega _{p}=1-\sum_{n\neq m}\frac{2e^{-\frac{1}{2}w_{nm}^{T}\Gamma
_{p}w_{nm}}}{\left\vert n-m\right\vert ^{3}}\left\langle
P_{nm}c_{n}^{\dagger }c_{m}\right\rangle w_{nm}w_{nm}^{T}
\end{equation}%
and $w_{l,nm}=\lambda _{ln}-\lambda _{lm}$, while for fermions the effective
mean-field matrix%
\begin{eqnarray}
\mathcal{F}_{f} &=&2i\sum_{n\neq m}\frac{e^{-\frac{1}{2}w_{nm}^{T}\Gamma
_{p}w_{nm}}}{\left\vert n-m\right\vert ^{3}}W_{f}^{\dagger }\frac{\delta }{%
\delta \Gamma _{m}}\left\langle P_{nm}c_{n}^{\dagger }c_{m}\right\rangle
W_{f}  \notag \\
&&+\left( 
\begin{array}{cc}
\mathcal{E}_{\mathrm{HF}} & \Delta _{\mathrm{F}} \\ 
\Delta _{\mathrm{F}}^{\dagger } & -\mathcal{E}_{\mathrm{HF}}^{T}%
\end{array}%
\right) +T\ln (\frac{1}{\Gamma _{f}}-1)
\end{eqnarray}%
is determined by%
\begin{eqnarray}
\mathcal{E}_{\mathrm{HF}} &=&[B+\frac{1}{2}V_{nn}+\sum_{n^{\prime
}}V_{nn^{\prime }}\left\langle c_{n^{\prime }}^{\dagger }c_{n^{\prime
}}\right\rangle -(\Delta _{x}^{T}G)_{n}]\delta _{nm}-V_{nm}\left\langle
c_{m}^{\dagger }c_{n}\right\rangle ,  \notag \\
\Delta _{\mathrm{F}} &=&V_{nm}\left\langle c_{m}c_{n}\right\rangle
,V_{nm}=2[(\lambda ^{T}\mathcal{K}\lambda )_{nm}+2F_{z}(\lambda
_{nm}+\lambda _{mn})].
\end{eqnarray}

The average value%
\begin{equation}
\left\langle P_{nm}c_{n}^{\dagger }c_{m}\right\rangle =-\frac{1}{4}%
\left\langle P_{nm}\right\rangle [\left( 1,i\right) \mathcal{S}\Theta \left( 
\begin{array}{c}
1 \\ 
-i%
\end{array}%
\right) ]_{mn}
\end{equation}%
of the string operator on the Gaussian state is determined by%
\begin{eqnarray}
\left\langle P_{nm}\right\rangle &=&(-1)^{N}s_{f}\text{Pf}(\frac{\Gamma _{F}%
}{2}),  \notag \\
\Gamma _{F} &=&\sqrt{1-\Theta }\Gamma _{m}\sqrt{1-\Theta }-i\sigma
^{y}(1+\Theta ),  \notag \\
\mathcal{S} &=&(i\sigma ^{y}\Gamma _{m}-1)\mathcal{T},\Gamma
_{m}=i(W_{f}\Gamma _{f}W_{f}^{\dagger }-1),  \notag \\
\mathcal{T} &=&\frac{1}{1+\frac{1}{2}(1-\Theta )(i\sigma ^{y}\Gamma _{m}-1)},
\notag \\
\Theta &=&I_{2}\otimes e^{i\pi \mathcal{N}},(\mathcal{N}_{l\notin \mathcal{S}%
_{nm}}=0,\mathcal{N}_{l\in \mathcal{S}_{nm}}=1),
\end{eqnarray}%
where $s_{f}=(-1)^{N/2}$ and $(-1)^{(N-1)/2}$ for the system with even and
odd modes, respectively. The functional derivatives are%
\begin{equation}
\frac{\delta }{\delta \Gamma _{m,ij}}\text{Pf}(\frac{\Gamma _{F}}{2})=-\frac{%
1}{2}\text{Pf}(\frac{\Gamma _{F}}{2})(\sqrt{1-\Theta }\frac{1}{\Gamma _{F}}%
\sqrt{1-\Theta })_{ij},
\end{equation}%
and%
\begin{eqnarray}
\frac{\delta }{\delta \Gamma _{m,ij}}\left\langle P_{nm}c_{n}^{\dagger
}c_{m}\right\rangle &=&-\frac{1}{2}\left\langle P_{nm}c_{n}^{\dagger
}c_{m}\right\rangle (\sqrt{1-\Theta }\frac{1}{\Gamma _{F}}\sqrt{1-\Theta }%
)_{ij}  \notag \\
&&+i\frac{1}{4}\left\langle P_{nm}\right\rangle [\mathcal{T}\Theta \left( 
\begin{array}{c}
1 \\ 
-i%
\end{array}%
\right) ]_{jn}[\left( 1,i\right) \mathcal{T}^{T}]_{mi}.
\end{eqnarray}

The free energy%
\begin{eqnarray}
f &=&\frac{1}{4}tr(\mathcal{K}\Gamma _{b,x}+\Gamma _{b,p})+\frac{1}{4}\Delta
_{x}^{T}\mathcal{K}\Delta _{x}+\sum_{n\neq m}\frac{e^{-\frac{1}{2}%
w_{nm}^{T}\Gamma _{p}w_{nm}}}{\left\vert n-m\right\vert ^{3}}\left\langle
P_{nm}c_{n}^{\dagger }c_{m}\right\rangle  \notag \\
&&+\sum_{n}[B+\frac{1}{2}V_{nn}-(\Delta _{x}^{T}G)_{n}+\frac{1}{2}%
\sum_{m}V_{nm}\left\langle c_{m}^{\dagger }c_{m}\right\rangle ]\left\langle
c_{n}^{\dagger }c_{n}\right\rangle  \notag \\
&&-\frac{1}{2}\sum_{nm}V_{nm}\left\langle c_{m}^{\dagger }c_{n}\right\rangle
\left\langle c_{n}^{\dagger }c_{m}\right\rangle +\frac{1}{2}%
\sum_{nm}V_{nm}\left\langle c_{m}^{\dagger }c_{n}^{\dagger }\right\rangle
\left\langle c_{n}c_{m}\right\rangle  \notag \\
&&-\sum_{j}\frac{d_{j}^{b}}{e^{\beta d_{j}^{b}}-1}+T\sum_{j}\ln (1-e^{-\beta
d_{j}^{b}})-\sum_{j}\frac{d_{j}^{f}}{e^{\beta d_{j}^{f}}+1}-T\sum_{j}\ln
(1+e^{-\beta d_{j}^{f}})
\end{eqnarray}%
monotonically decreases in the imaginary time evolution, where $d_{j}^{b}$
and $d_{j}^{f}$ are the positive eigenvalues of $T\sigma ^{y}\ln \frac{%
\Gamma _{b}\sigma ^{y}+1}{\Gamma _{b}\sigma ^{y}-1}$ and $-T\ln (\frac{1}{%
\Gamma _{f}}-1)$, respectively.

\begin{figure*}[t!]
\centering 
\includegraphics[width=0.72\textwidth]{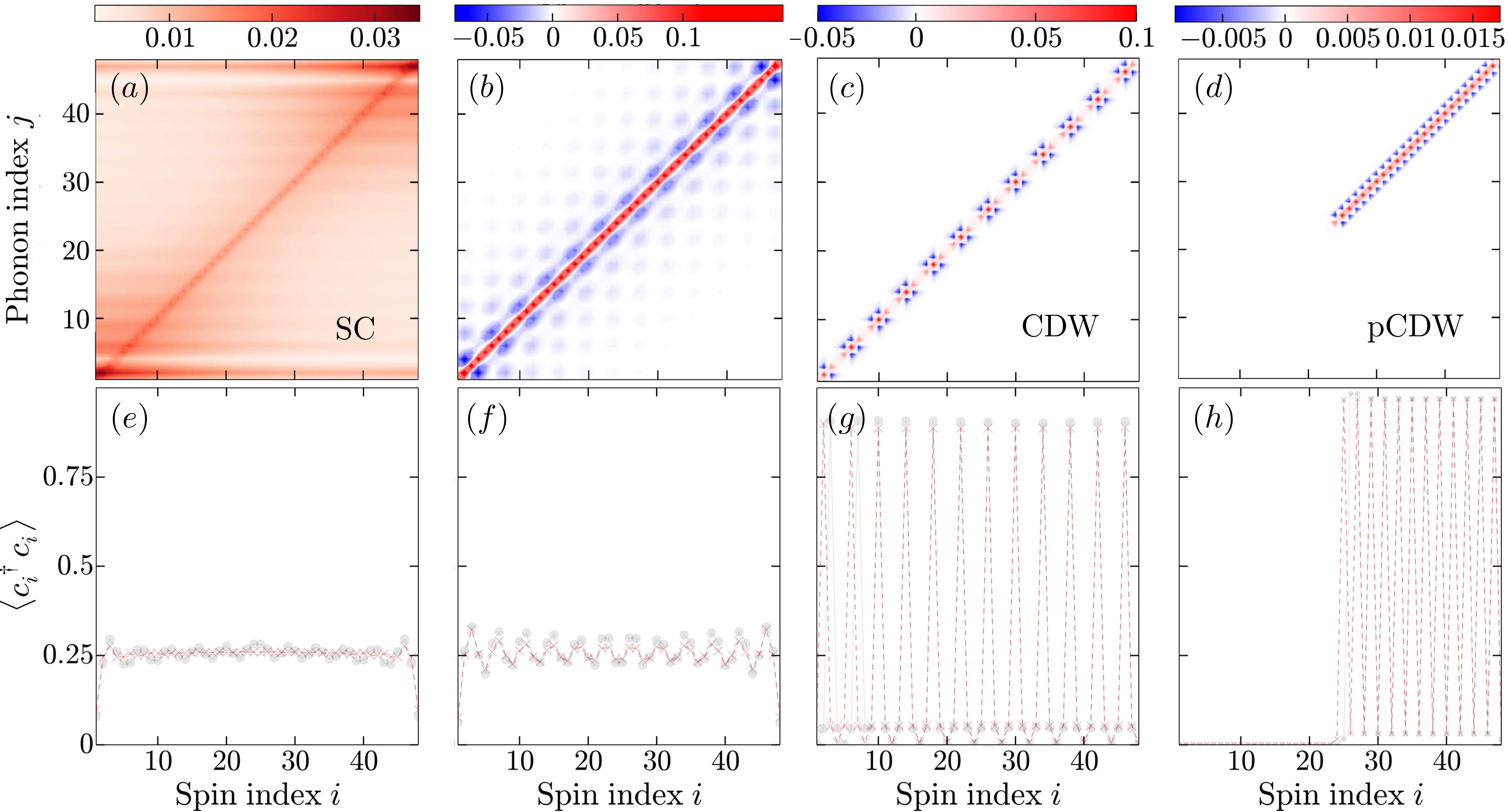}
\caption{\label{fig:spin-phonon-app}Spin phonon correlator $\Pi_{ij}$ and spin configuration for different ground states at quarter filling $\nu = 1/4$ as obtained with NGS.
\textit{Upper panel}: Spin-phonon correlator $\Pi$ as defined in Eq.~\eqref{eq:spin-phonon-ngs}.
\textit{Lower panel}: Site-dependent density $\langle c_i^\dagger c_i \rangle$.
The grey dots indicate the DMRG result.
The four columns correspond to particular choices for $F_z$ and $\nu_\beta = \log \beta$ (compare right panel of Fig.~2).
(\textit{a}) and (\textit{e}): $F_z = 1.6$, $\nu_\beta = 2.1$ (SC regime).
(\textit{b}) and (\textit{f}): $F_z = 0.6$, $\nu_\beta = -2.1$ (precursor of CDW regime).
(\textit{c}) and (\textit{g}): $F_z = 1.2$, $\nu_\beta = -2.1$ (CDW regime).
(\textit{d}) and (\textit{h}): $F_z = 1.5$, $\nu_\beta = -2.1$ (pCDW regime).
Other numerical parameters: equivalent to Fig.~3 in the main text.
}
\end{figure*}

To characterize the SC, CDW, and PS phases, we calculate the displacement%
\begin{equation}
\left\langle r_{l}\right\rangle =\Delta _{x,l}-2\sum_{n}\lambda
_{ln}\left\langle c_{n}^{\dagger }c_{n}\right\rangle ,
\end{equation}%
of phonons, and the order parameters%
\begin{eqnarray}
O_{\mathrm{CDW}} &=&\frac{1}{N}\sum_{n}(-1)^{n}\left\langle c_{n}^{\dagger
}c_{n}\right\rangle ,  \notag \\
O_{\mathrm{SC}} &=&\left\langle \sigma _{m}^{-}\sigma _{n}^{-}\right\rangle .
\end{eqnarray}%
The SC order parameter%
\begin{equation}
O_{\mathrm{SC}}=e^{-\frac{1}{2}\bar{w}_{nm}^{T}\Gamma _{p}\bar{w}%
_{nm}}\left\langle P_{nm}c_{m}c_{n}\right\rangle sgn(m-n),
\end{equation}%
is determined by the phonon dressing factor $\bar{w}_{l,nm}=\lambda
_{ln}+\lambda _{lm}$ and the average value%
\begin{equation}
\left\langle P_{nm}c_{m}c_{n}\right\rangle =-\frac{1}{4}\left\langle
P_{nm}\right\rangle [(1,i)\mathcal{S}\left( 
\begin{array}{c}
1 \\ 
i%
\end{array}%
\right) ]_{nm}.
\end{equation}%
The connected correlation function $\Pi _{ln}=\left\langle \bar{r}%
_{l}c_{n}^{\dagger }c_{n}\right\rangle -\left\langle \bar{r}%
_{l}\right\rangle \left\langle c_{n}^{\dagger }c_{n}\right\rangle \equiv
\left\langle \bar{r}_{l}c_{n}^{\dagger }c_{n}\right\rangle _{c}$ between the
spin at the site $n$ and the phonon at the site $l$ reads%
\begin{equation}\label{eq:spin-phonon-ngs}
\Pi _{ln}=-2\sum_{m}\lambda _{lm}D_{mn}.
\end{equation}

In Fig.~\ref{fig:spin-phonon-app}, we show the numerical results obtained with the NGS ansatz for the same numerical parameters used to obtain Fig.~3 in the main text.
We find that the results agree well, both qualitatively and quantitatively.
The only quantitatively different result concerns the SC regime.
Here we find that the NGS ansatz overestimates the spin-phonon correlation, cf.~Fig.~\ref{fig:spin-phonon-app}(a).
The stripe pattern from Fig.~3(a) is still present, but less pronounced.

To characterize the phonon excitation number, we expand $\bar{r}%
_{l}=\sum_{k}\mathcal{M}_{lk}(a_{k}^{\dagger }+a_{k})/\sqrt{\Omega _{k}}$
and $\bar{p}_{l}=i\sum_{k}\mathcal{M}_{lk}\sqrt{\Omega _{k}}(a_{k}^{\dagger
}-a_{k})$ in terms of normal modes for the non-interacting phonon. The
density of coherent phonons is%
\begin{equation}
n_{c}=\frac{1}{N}\sum_{k}\left\vert \left\langle a_{k}\right\rangle
\right\vert ^{2}=\frac{1}{4N}\sum_{k}\Omega _{k}(\mathcal{M}%
_{kl}^{T}\left\langle \bar{r}_{l}\right\rangle )^{2}.
\end{equation}%
The quantum fluctuation of the phonon density is characterized by%
\begin{eqnarray}
n_{s} &=&\frac{1}{N}\sum_{k}\left\langle a_{k}^{\dagger }a_{k}\right\rangle
-n_{c}  \nonumber \\
&=&\frac{1}{4N}tr[\Omega \mathcal{M}^{T}(\Gamma _{b,x}+4\lambda D\lambda ^{T}%
\mathcal{)M}+\frac{1}{\Omega }\mathcal{M}^{T}\Gamma _{b,p}\mathcal{M]}\text{.%
}
\end{eqnarray}

\begin{figure}[b!]
\includegraphics[width=0.95\columnwidth]{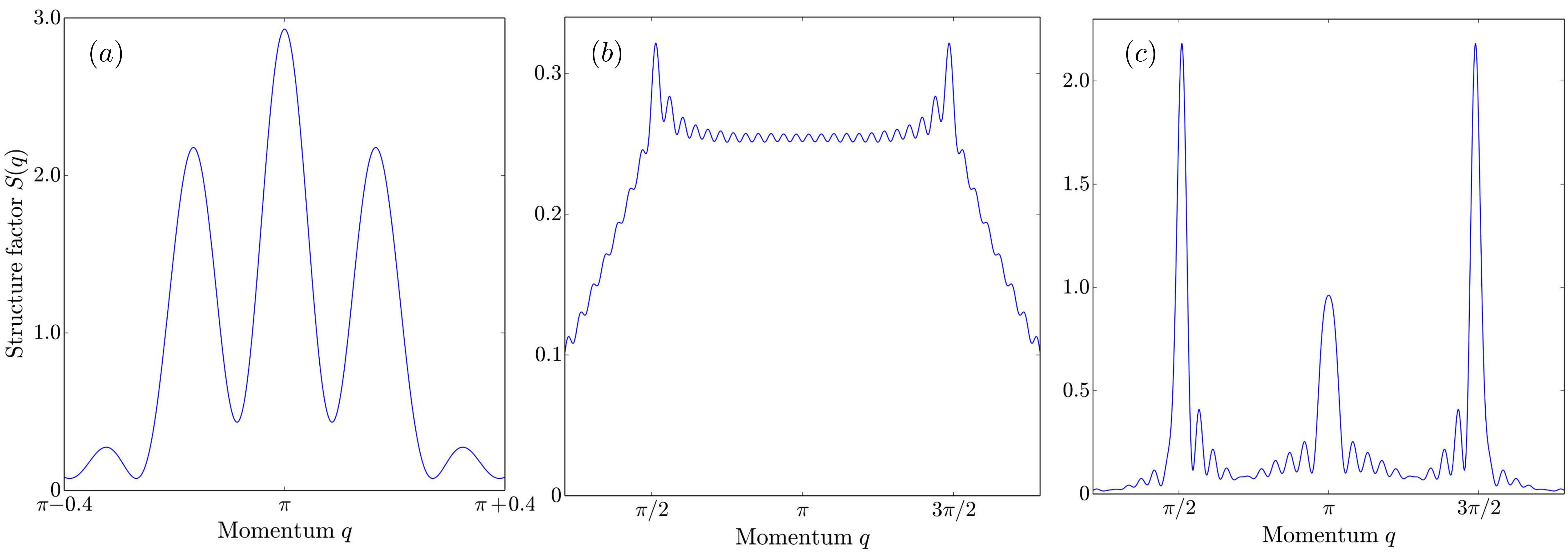}
\caption{\label{fig:structure_SM}Structure factor $S(q)$.
(\textit{a}) Half filling $\nu = 1/2$, $\beta = -2.4$, $F_z = 2.1$,
(\textit{b}) quarter filling $\nu = 1/4$, $\beta = -2.1, F_z = 0.6$,
(\textit{c}) quarter filling $\nu = 1/4$, $\beta = -2.1, F_z = 1.2$.
Other numerical parameters: $N = 48$, $\omega_z/J = 1$.
}
\end{figure}

\section{Other observables \label{sm:other-observables}}

Here we discuss additional observables that are not shown in the main text.

\subsection{Structure factor \label{sm:structure-factor}}

We study the structure factor via the spin-spin correlations as
\begin{equation}
    S(q) = \frac{1}{N} \sum_{i,j} \langle \sigma_i^z \sigma_j^z \rangle e^{iq |i-j|},
\end{equation}
and show results for the CDW state in Fig.~\ref{fig:structure_SM} that correspond to the cases studied in the main text.
At half filling ($\nu = 1/2$), the structure factor displays a peak at $q = \pi$ as expected, and shows two additional peaks nearby.
Figs.~\ref{fig:structure_SM}(\textit{b}) and (\textit{c}) show the structure factor in the stiff limit ($\beta \ll 1$) at quarter filling $\nu = 1/4$.
At a comparatively small coupling $F_z = 0.6$ (compare Figs.~3(\textit{b}) and (\textit{f}) from the main text), two peaks start to evolve at $q = \pi/2$ and $q = 3\pi/2$.
Only when $F_z$ is increased, $S(q)$ features two prominent peaks at $q = \pi/2$ and $q = 3\pi/2$, and another peak at $q = \pi$.
The latter is related to the non-vanishing background of the CDW state with period $4$ shown in Fig.~2(\textit{g}).

\subsection{Phonon observables \label{sm:phonons}}

We study the staggered phonon parameter
\begin{equation}
    m_\mathrm{ph} = \frac{1}{N} \sum_{n=1}^N (-1)^n \langle b_n + b_n^\dagger \rangle.
\end{equation}
In the soft limit ($\beta \gg 1$), we find that $m_\mathrm{ph}$ suddenly increases from zero to a finite value at the phase transition from SC to PS.
This is a contribution from the domain wall in the phase-separated regime and approaches a constant finite value as $F_z$ is increased and the width of the domain wall tends to zero.

\subsection{Width of domain wall \label{sm:width-ps}}

\begin{figure}[t!]
\includegraphics[width=0.55\columnwidth]{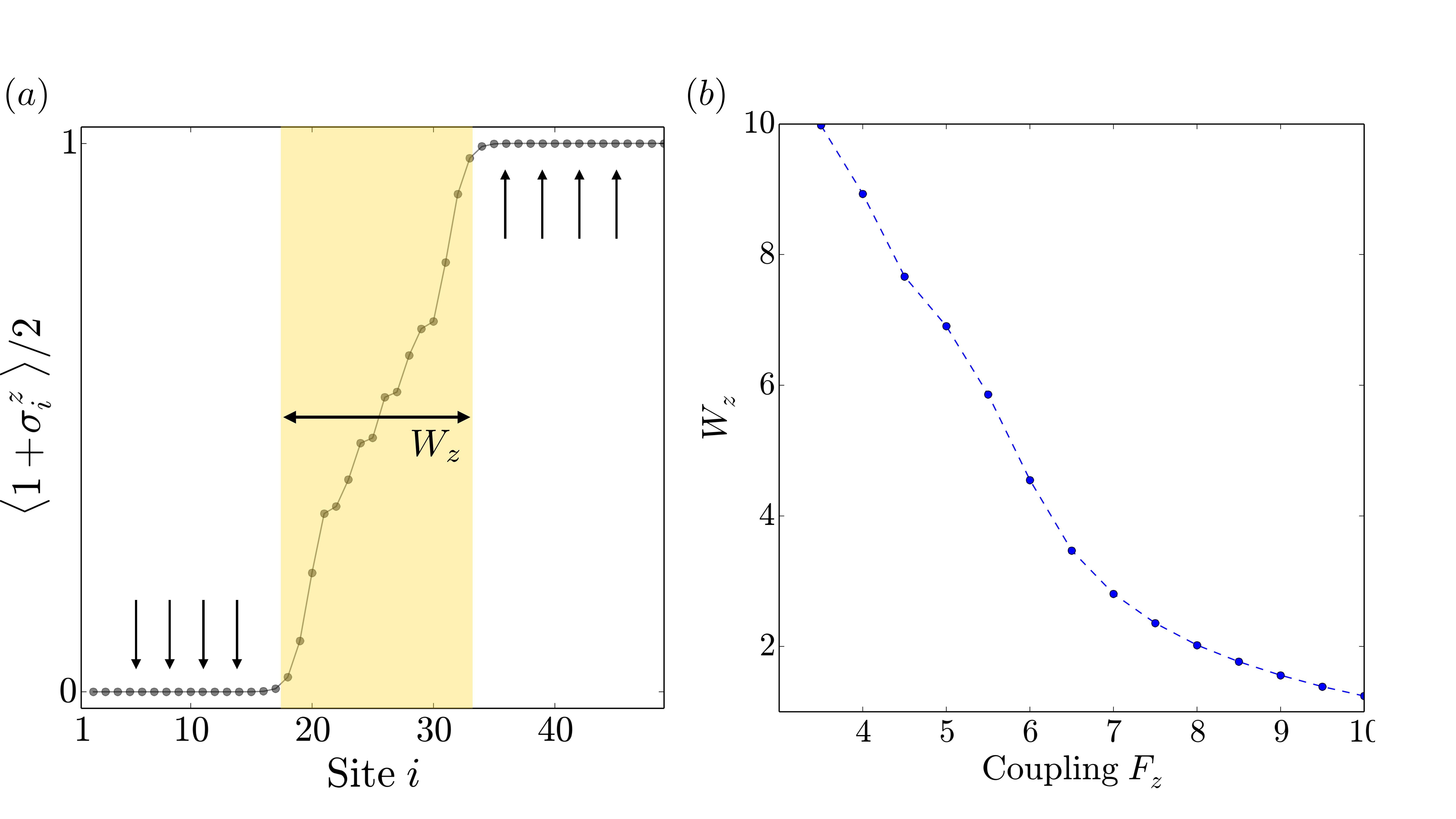}
\caption{\label{fig:width_SM}Width of domain wall $W_z$ in the case of phase separation.
(\textit{a}) Sketch of domain wall and width $W_z$ at half filling $\nu = 1/2$.
(\textit{b}) Width $W_z$ as a function of coupling $F_z$ in soft limit at $\nu_\beta = 3$ and at $\nu = 1/2$.
}
\end{figure}

In the case of phase separation, the domain wall separating the two phases shrinks as $F_z$ increases.
We calculate the width of the domain wall and define
\begin{equation}\label{eq:SM_width_domain_wall}
W_z = \sum_{n=1}^N \frac{(z_n - z_\mathrm{c})^2}{2} \left ( 1 - | 1 - | \langle \sigma_n^z \rangle - 1 | | \right ),
\end{equation}
where $z_n$ denotes the $n$th ion position and $z_\mathrm{c}$ is the center of the domain wall.
An exemplary result is shown in Fig.~\ref{fig:width_SM}.

\end{document}